\documentclass[10pt,onecolumn,letterpaper]{article}

\usepackage[pagenumbers]{cvpr} 

\definecolor{cvprblue}{rgb}{0.21,0.49,0.74}
\usepackage[pagebackref,breaklinks,colorlinks,allcolors=cvprblue]{hyperref}

\def\confName{COMPSCI690@UMass}
\def\confYear{2025}

\title{SCOPE: Siamese Contrastive Operon Pair Embeddings for Functional Sequence Representation and Classification}

\author{Akarsh Gupta\\
{\tt\small akarshgupta@umass.edu}
\and
Kenneth Rodrigues\\
{\tt\small kerodrigues@umass.edu}
\and
Sagnik Chatterjee\\
{\tt\small sagnikchatte@umass.edu}
}

\begin{document}
\maketitle
\begin{abstract}
Identifying operons is a fundamental step in understanding prokaryotic gene regulation, 
as the classification of genes into operons contributes directly to the reconstruction 
of regulatory networks, functional annotation of unannotated genes, and the development 
of drug candidates \cite{MorenoHagelsieb2015, Okuda2011}. While experimental approaches such as RT-PCR and RNA-seq provide precise evidence 
of operon structure, they are laborious and largely limited to well-studied model 
organisms, making scalable computational methods essential for genome-wide operon 
identification \cite{Osmanbeyoglu2010}. Existing computational approaches to operon prediction have employed traditional 
machine learning classifiers, including logistic regression \cite{Westover2005} 
and decision tree-based methods \cite{Krishnakumar2022}, motivating our use of 
these as physicochemical baselines in our experiments. Additonally, the DGEB benchmark evaluates operonic pair classification by embedding each sequence 
independently using a pre-trained protein language model and computing pairwise cosine 
similarity \cite{WestRoberts2024}. In contrast, our Siamese MLP learns a 
classifier over the fused embedding space, which is 
theoretically better motivated for binary classification as cosine similarity 
has been shown to yield potentially meaningless similarity scores 
depending on the regularization of the embedding model \cite{Steck2024}. While protein language model embeddings substantially outperform physicochemical 
features in terms of ROC-AUC, we find that a learned Siamese MLP classification 
head does not significantly improve over unsupervised cosine similarity in Average 
Precision for operonic pair classification, suggesting that the geometry of the 
embedding space already captures the functional relationships necessary for this task. However, our Siamese MLP achieves a ROC-AUC of 0.71, competitive with 
state-of-the-art models on the DGEB leaderboard, demonstrating that 
this is a
promising direction for operonic pair classification. These findings suggest that protein language model embeddings 
are a viable and scalable foundation for operonic pair classification across 
diverse microbial genomes, with direct implications for automated genome 
annotation, regulatory network reconstruction, and the functional characterization 
of organisms whose genomes lack experimental operon annotations. Code is available at \footnote{\url{https://github.com/kennethnrk/Operon-pair-classification}}.

\end{abstract}

\section{Introduction}

Understanding the functional relationships between proteins is a central challenge 
in computational biology \cite{MorenoHagelsieb2015, Okuda2011}. 
Pair classification, the task of determining whether two biological sequences share 
a functional relationship, is one such challenge that directly probes whether learned 
sequence representations encode biologically meaningful information \cite{WestRoberts2024}.
Operon pair classification specifically refers to the binary classification task that evaluates whether a consecutive pair of genomic sequences belongs to the same 
transcription unit (operon) to measure a model's understanding of their 
functional relationship \cite{WestRoberts2024}. Given the traditionally perceived importance of operons in co-regulating genes whose products functionally interact, they have been central in the field of comparative genomics aiming at predicting functional associations \cite{MorenoHagelsieb2015}.

This project evaluates the performance of Siamese embedding-based encoders coupled with neural networks in operon classification, comparing their utility against traditional logistic regression and XGBoost baselines, arranged in a Siamese format. 
Our logistic regression and XGBoost models leverage physicochemical features of the proteins, extracted from the amino-acid sequence data, representing the traditional models in our experiment. Our encoder-based architecture models leverage embeddings generated by foundational models that are further processed by a multi-layer perceptron neural network (MLP). Our architectural choices are motivated based on the results of embedding-based models such as ESM-2 \cite{Lin2023} and ProtBERT. 

We ask whether embedding-based encoders, when paired with a Siamese MLP 
architecture, better capture the functional relationships required for operon 
pair classification than traditional classifiers operating on hand-crafted 
physicochemical features. We hypothesize that the richer sequence representations 
produced by pre-trained protein language models will yield superior classification 
performance, as physicochemical features alone may fail to encode the higher-order 
contextual dependencies that distinguish co-operonic from non-co-operonic protein pairs.
\cite{Elnaggar2021} demonstrating strong capability in capturing 
the functional information encoded in protein sequences.

\section{Method}
Our training and validation data are sourced from ODB, a database of known and conserved operons across sequenced microbial genomes \cite{Okuda2011}. This dataset consists of pairs of amino acid sequences along with a label indicating whether the proteins are operonic or not. For our final evaluation, we've used data from the DGEB, implicitly pairing neighboring proteins.

\subsection{Baseline models}

Our baseline implementations consist of Siamese logistic regression and XGBoost 
models. For XGBoost, we use \texttt{XGB\allowbreak Classifier} from the open-source 
\texttt{xgboost} Python package. For logistic regression, we use 
\texttt{LogisticRegression} from \texttt{scikit-learn}'s \texttt{linear\_model} module. 

Our input data consists of physicochemical properties of the proteins represented using a 305-dimensional feature vector from each protein pair by computing 
per-sequence physicochemical statistics, including amino acid composition, 
biochemical group frequencies, hydrophobicity, net charge, molecular weight, 
and Shannon entropy, and combining them via a Siamese interaction pattern 
of concatenation, signed difference, absolute difference, and element-wise product. Both models output a binary label (0 or 1), indicating whether a pair of sequences is operonic, this is evaluated against the original label in the dataset, which is also binary.

We train logistic regression with $C=1.0$, an L2 penalty, and \texttt{class\_weight="balanced"} 
to account for class imbalance, using the \texttt{lbfgs} solver with a maximum of 1000 iterations. 
Input features are standardized via \texttt{StandardScaler} prior to training, as logistic regression 
is sensitive to feature scale. We train XGBoost with 300 estimators, a maximum tree depth of 6, and a learning rate of 0.05, 
with \texttt{scale\_pos\_weight} set to the negative-to-positive class ratio to handle class imbalance. 
Early stopping with a patience of 20 rounds is applied on the validation AUC to prevent overfitting. Both models were trained using a purely CPU-based computational environment provided by Kaggle.
\subsection{Encoder based models}
Our encoder-based architecture consists of Siamese transformer models that generate 
embeddings for each sequence, which are then mean-pooled to perform dimensionality 
reduction. The resulting embedding vectors for each sequence in the pair are then 
fused via concatenation, signed difference, absolute difference, and element-wise 
product, mirroring the Siamese interaction pattern used in our physicochemical 
baselines, before being passed to an MLP that ultimately classifies the pair 
as operonic or not. Our input data consists of the raw amino-acid sequences, and the output consists of a binary label similar to the baseline models.

We employ two encoder models with distinct architectures. ESM-2 3B \cite{Lin2023} 
is a transformer-based protein language model trained on 250 million protein sequences 
from UniRef50, scaling to 3 billion parameters, and has demonstrated strong performance 
on a wide range of protein function prediction tasks. ProtBERT-BFD \cite{Elnaggar2021} 
is a BERT-based protein language model trained on the Big Fantastic Database (BFD), 
comprising over 2.1 billion protein sequences, and has been shown to capture important 
biophysical properties of proteins directly from sequence data. Both models are used 
as frozen encoders, with their weights kept fixed during training, such that only 
the downstream MLP is trained on the operon pair classification task. Both encoder models are loaded and used via the Hugging Face \texttt{transformers} 
library, while the MLP classifier is implemented in PyTorch using \texttt{torch.nn.Sequential}, 
composed of \texttt{torch.nn.Linear} layers with \texttt{torch.nn.ReLU} activations 
and a final \texttt{torch.nn.Sigmoid} output layer.

Both encoder models accept variable-length amino acid sequences as input, 
producing per-token embeddings that are mean-pooled across the sequence 
length to yield a fixed-size representation. ESM-2 3B produces a 
2560-dimensional embedding per sequence, while ProtBERT-BFD produces a 
1024-dimensional embedding per sequence. Following mean-pooling, the 
embeddings for each sequence pair are fused via the Siamese interaction 
pattern (concatenation, signed difference, absolute difference, and 
element-wise product), yielding a fused vector of dimension $5 \times 2560 = 12800$ 
for ESM-2 3B and $5 \times 1024 = 5120$ for ProtBERT-BFD. This fused 
vector is passed to the MLP classifier, which maps it to a single scalar 
output via a sigmoid activation, producing a probability in $[0, 1]$ that 
the pair belongs to the same transcription unit, which is thresholded at 
0.5 to produce the final binary prediction $\hat{y} \in \{0, 1\}$. Embeddings were generated and the MLP classifier was trained on Kaggle's cloud 
computational environment, utilizing a single NVIDIA T4 GPU with 16GB of memory.

The MLP hyperparameters were selected empirically based on validation AUROC 
performance. For the ESM-2 3B model, the MLP consists of four hidden layers 
with dimensions $[4096, 1024, 256, 64]$, a dropout rate of 0.5, and a batch 
size of 256, trained for up to 200 epochs with early stopping applied after 
20 epochs of no improvement in validation AUROC. For the ProtBERT-BFD model, 
the MLP consists of four hidden layers with dimensions $[2048, 512, 128, 32]$, 
a dropout rate of 0.3, and a batch size of 128, trained for up to 50 epochs. 
Both models are optimized using AdamW with a learning rate of $1 \times 10^{-4}$, 
with weight decay set to 0.1 for ESM-2 3B and $1 \times 10^{-2}$ for ProtBERT-BFD. 
A cosine annealing learning rate scheduler is applied in both cases. Class imbalance 
is addressed by setting \texttt{pos\_weight} in \texttt{BCEWithLogitsLoss} to the 
negative-to-positive class ratio, and label smoothing of 0.1 is applied to reduce 
overconfidence. Gradient clipping with a maximum norm of 1.0 is used during training 
to stabilize optimization.
\begin{figure}[h]
    \centering
    \includegraphics[width=\linewidth]{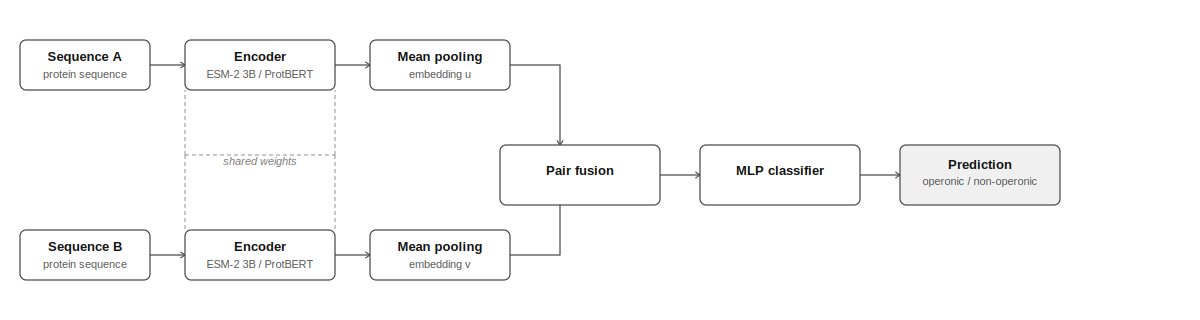}
    \caption{Architecture of the Siamese encoder model for operon pair classification. Both sequences are processed by a shared-weight encoder, mean-pooled into fixed-size embeddings, fused via concatenation and interaction operations, and classified by an MLP.}
    \label{fig:architecture}
\end{figure}

\section{Results and Conclusion}
The primary metric used in DGEB for operon pair classification is the precision \cite{WestRoberts2024}. We evaluate all models using Accuracy, Precision, Recall, F1 Score, and ROC-AUC. 
Accuracy measures the fraction of correctly classified pairs, while Precision and Recall 
capture the trade-off between false positives and false negatives respectively,
particularly important given the class imbalance inherent in operon pair datasets. 
F1 Score, the mean of Precision and Recall, provides a single balanced metric 
that penalizes models that sacrifice one for the other.  

\begin{table}[h]
    \centering
    \caption{Average Precision (AP) across models on the operon pair classification task.}
    \label{tab:ap_results}
    \begin{tabular}{lcc}
        \hline
        \textbf{Model} & \textbf{Type} & \textbf{Average Precision} \\
        \hline
        Logistic Regression & Physicochemical Baseline & 0.4100 \\
        XGBoost             & Physicochemical Baseline & 0.4000 \\
        \hline
        ESM2-3B + MLP       & Siamese MLP (Ours)       & 0.4500 \\
        ProtBERT-BFD + MLP  & Siamese MLP (Ours)       & 0.5074 \\
        \hline
        DGEB Baseline       & Reference                & 0.5247 \\
        \hline
    \end{tabular}
\end{table}

The operon pair classification datasets in DGEB exhibit significant class 
imbalance, as the majority of consecutive gene pairs in a genome do not 
belong to the same transcription unit \cite{WestRoberts2024, MorenoHagelsieb2015}, 
ROC-AUC summarizes model 
performance across all classification thresholds, where 0.5 represents random chance 
and 1.0 represents perfect discrimination, making it robust to class imbalance and a suitable evaluation metric. \cite{Richardson2024}. 
\begin{figure}[h]
    \centering
    \includegraphics[width=\linewidth]{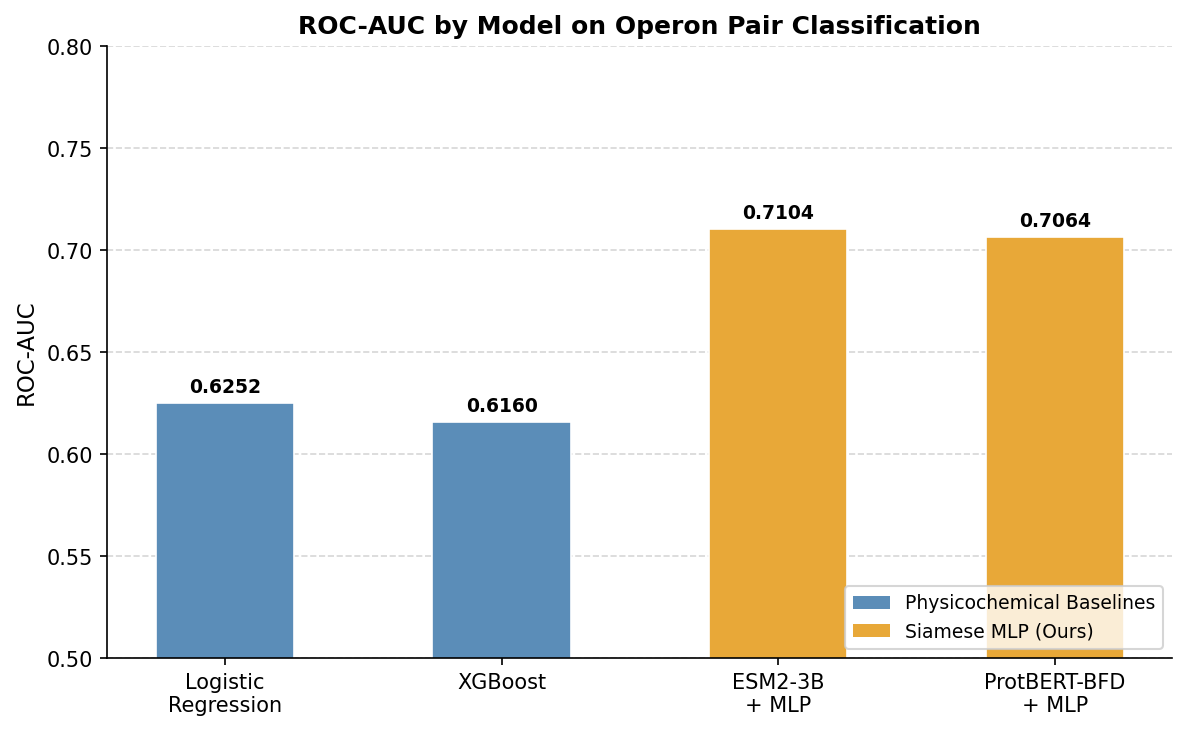}
    \caption{ROC-AUC across models on the operon pair classification task. Blue bars represent physicochemical baselines and orange bars represent our Siamese MLP models. DGEB does not report a ROC-AUC baseline.}
    \label{fig:auc_comparison}
\end{figure}

Table \ref{tab:ap_results} and Figure \ref{fig:auc_comparison} summarize the 
performance of all models on the operon pair classification task. The physicochemical 
baselines perform modestly, with logistic regression and XGBoost achieving ROC-AUC 
scores of 0.6252 and 0.6160, and Average Precision of 0.41 and 0.40 respectively, 
suggesting that hand-crafted physicochemical features alone are insufficient to fully 
capture the functional relationships required for operon pair classification. Our 
Siamese MLP models coupled with pre-trained protein language model encoders yield 
meaningful improvements, with ESM-2 3B and ProtBERT-BFD achieving ROC-AUC scores 
of 0.7104 and 0.7064, and Average Precision of 0.5172 and 0.5074 respectively, 
representing an improvement of approximately 10 percentage points in ROC-AUC over 
the physicochemical baselines. Comparing against the DGEB baseline, which employs 
ESM-2 embeddings with cosine similarity and achieves an Average Precision of 0.5247, 
our ESM-2 3B Siamese MLP model performs comparably at 0.5172, falling just short of 
the DGEB baseline despite using a more complex classification head. This suggests 
that the representational quality of the ESM-2 embeddings is the primary driver of 
performance, and that the additional complexity of the Siamese MLP architecture does 
not yield significant gains over a simple similarity-based approach on this task.

Notably, despite the marginal gap in Average Precision relative to the DGEB baseline, 
our ESM-2 3B Siamese MLP achieves a ROC-AUC of 0.7104, which is comparable to the 
accuracy reported on the DGEB leaderboard for ESM-3 \cite{WestRoberts2024}, a 
significantly newer and larger model. This indicates that our Siamese MLP architecture 
is able to extract competitive discriminative signal from an older encoder, suggesting 
that the architectural choice of learned pairwise fusion may compensate for the 
representational limitations of smaller, earlier-generation protein language models, 
and warrants further exploration with more recent encoders.

\section{Conclusion}
Our results support the hypothesis that embedding-based models better capture 
the functional relationships required for operon pair classification than 
physicochemical features, with our Siamese MLP models outperforming both 
baselines by approximately 10 percentage points in ROC-AUC. However, the 
marginal gap in Average Precision relative to the DGEB cosine similarity 
baseline suggests that the learned classification head does not add significant 
value over direct embedding similarity for this task, implying that the 
representational quality of the encoder is the primary driver of performance. 
Embedding-based models are clearly justified in this space, and future work 
with larger or more recent encoders, or alternative fusion strategies beyond 
element-wise interaction, may close the remaining gap to state-of-the-art.
{
    \small
    \bibliographystyle{ieeenat_fullname}
    \bibliography{main}
    
}

\end{document}